\documentclass[preprint,showpacs,preprintnumbers,superscriptaddress,amsmath,amssymb, pre]{revtex4}

\makeatletter

\usepackage{graphicx}
\usepackage{cases}
\usepackage{dcolumn}
\usepackage{bm}
\usepackage{color}

\begin{document}

\preprint{PRE}

\title{Individual position diversity in dependence socioeconomic network increases economic output}

\author{Wen-Jie Xie}
 \affiliation{School of Business, East China University of Science and Technology, Shanghai 200237, China} %
  \affiliation{Research Center for Econophysics, East China University of Science and Technology, Shanghai 200237, China} %

 \author{Yan-Hong Yang}
 \affiliation{School of Business, East China University of Science and Technology, Shanghai 200237, China} %
  \affiliation{Research Center for Econophysics, East China University of Science and Technology, Shanghai 200237, China} %

  \author{Ming-Xia Li}
   \affiliation{Research Center for Econophysics, East China University of Science and Technology, Shanghai 200237, China} %
 \affiliation{Research Institute of Sports Economics, East China University of Science and Technology, Shanghai 200237, China} %

\author{Zhi-Qiang Jiang}
 \affiliation{School of Business, East China University of Science and Technology, Shanghai 200237, China} %
 \affiliation{Research Center for Econophysics, East China University of Science and Technology, Shanghai 200237, China} %

\author{Wei-Xing Zhou}
 \email{wxzhou@ecust.edu.cn}
 \affiliation{School of Business, East China University of Science and Technology, Shanghai 200237, China} %
 \affiliation{Research Center for Econophysics, East China University of Science and Technology, Shanghai 200237, China} %
 \affiliation{Departmenent of Mathematics, East China University of Science and Technology, Shanghai 200237, China}

\date{\today}

\begin{abstract}
   {The} availability of big data recorded from massively multiplayer online role-playing game (MMORPG) allows us to gain a deeper understanding the potential connection between individuals' network positions and their economic outputs. We use a statistical filtering method to construct dependence networks from weighted friendship networks of individuals. We investigate the 30 distinct motif positions in the 13 directed triadic motifs which represent microscopic dependence among individuals. Based on the structural similarity of motif positions, we further classify individuals into different groups. The node position diversity of individuals is found to be positively correlated with their economic outputs. We also find that the economic outputs of leaf nodes are significantly lower than that of the other nodes in the same motif. Our findings shed light on understanding the influence of network structure on economic activities and outputs in social system.
\end{abstract}


\maketitle

\section{Introduction}
\label{S1:Introduction}

Considerable studies have offered us a deep understanding of the influence of network structures on the dynamics of complex systems, such as the spreading of diseases and information \cite{Wang-Gonzalez-Hidalgo-Barabasi-2009-Science} and emerging of collaborations \cite{Newman-2001-PNAS}. However, the connection between network position and economic output is less studied. It is reported that specific network structures may enhance economic outputs \cite{Bhattacharya-Dugar-2014-MS, Xie-Li-Jiang-Zhou-2014-SR, Eagle-Macy-Claxton-2010-Science, Bettencourt-Samaniego-Youn-2014-SR, Ortman-Cabaniss-Sturm-Bettencourt-2015-SciAdv}.
Eagle et al argue that the diversity of individual relationships within a community strongly correlates with economic development of communities \cite{Eagle-Macy-Claxton-2010-Science}. Furthermore, Bettencourt et al and Ortman et al  find that the diversity of relationships is positively correlated with the productivity of individuals and communities \cite{Bettencourt-Samaniego-Youn-2014-SR, Ortman-Cabaniss-Sturm-Bettencourt-2015-SciAdv}.


Network motifs are building blocks of complex networks \cite{Milo-Itzkovitz-Kashtan-Levitt-ShenOrr-Ayzenshtat-Sheffer-Alon-2004-Science, Milo-ShenOrr-Itzkovitz-Kashtan-Chklovskii-Alon-2002-Science, Milo-Kashtan-Itzkovitz-Newman-Alon-2004-XXX}. It is found that network motifs of social networks may reflect the driving forces for forming social structures \cite{Kovanen-Kaski-Kertesz-Saramaki-2013-PNAS,Klimek-Thurner-2013-NJP}. Similar to the friendship networks of US students \cite{Ball-Newman-2013-NS}, Xie et al. studied triadic motifs in dependence networks of virtual societies and found that low level individuals have preference of forming links to high level individuals \cite{Xie-Li-Jiang-Zhou-2014-SR}. Their findings in virtual world are consistent with empirical findings in real society that ``collaboration is easier when both partners share the same social status, and the probability of partnership formation decreases significantly as the status gap between the partners increases'' \cite{Bhattacharya-Dugar-2014-MS}.

Understanding the structure and function of social networks are of great importance to investigate economic activities and {{outputs}} in social systems \cite{Kovanen-Kaski-Kertesz-Saramaki-2013-PNAS, Palla-Barabasi-Vicsek-2007-Nature, Onnela-Saramaki-Hyvonen-Szabo-Lazer-Kaski-Kertesz-Barabasi-2007-PNAS, Kumpula-Onnela-Saramaki-Kaski-Kertesz-2007-PRL, Eagle-Penland-Lazer-2009-PNAS, Jo-Pan-Kaski-2011-PLoS1, Jiang-Xie-Li-Podobnik-Zhou-Stanley-2013-PNAS}. In real world, social networks are not large and samples are biased \cite{Ball-Newman-2013-NS,Currarini-Jackson-Pin-2009-Em,Currarini-Jackson-Pin-2010-PNAS}, which hinders the empirical investigation of relationship between network structures and economic activities in social system.
Some other empirical studies have shown that individuals' or firms' network positions are closely related to success measured in terms of economic production \cite{Uzzi-1996-ASR,Guimera-Uzzi-Spiro-Amaral-2005-Science,Cantner-Joel-2011-IUPJKM,Garas-Tomasello-Schweitzer-2014-ArXiv}.
In the era of big data, information technology provides us alternative methods to collect data of social relationships and economic activities, for example, massively multiplayer online role-playing games (MMORPGs) \cite{Jiang-Zhou-Tan-2009-EPL,Jiang-Ren-Gu-Tan-Zhou-2010-PA,Thurner-Szell-Sinatra-2012-PLoS1,Szell-Sinatra-Petri-Thurner-Latora-2012-SR,Szell-Thurner-2012-ACS}, which enable us to study complex social and economic behaviours of human in online social systems \cite{Bainbridge-2007-Science,Papagiannidis-Bourlakis-Li-2008-TFSC,Williams-2010-CT}. Bainbridge et al also emphasized the scientific potential of virtual world for future research \cite{Bainbridge-2007-Science}. Empirical studies show that social behaviours in MMORPGs are representative of human behaviours in many aspects in real society \cite{Chesney-Chuah-Hoffmann-2009-JEBO, Szell-Lambiotte-Thurner-2010-PNAS, Szell-Thurner-2010-SN, Klimek-Thurner-2013-NJP, Szell-Thurner-2013-SR, Grabowski-Kosinski-2008-APPA, Grabowski-Kruszewska-2007-IJMPC}.

In this paper, we construct dependence networks based on weighted friendship networks of individuals and identify 30 distinct motif positions in 13 directed triadic motifs which represent local dependence among individuals. Using the $k$-means algorithm, we further classify individuals into $k$ clusters based on the motif position profiles. Our results indicate that the motif position of individuals do have great influence on their economic outputs.

\section{Materials and Methods}
\label{section:Materials and Methods}

\subsection{Data description}
We use a huge database recorded from $124$ servers to investigate the potential connection between network structure and economic output for individuals within the virtual world of a popular Massively Multiplayer Online Role-Playing Game (MMORPG) in China. There are two opposing camps or societies in a virtual world residing in a server, thus giving us $248$ virtual societies. There exist great differences about the numbers of avatars among different virtual societies. The populations of virtual worlds vary from thousands up to fifty thousands. The distribution of the number of avatars in each virtual world is drawn in Fig. 1G. In each society, three professions have different skills. The advantage of warrior's skill is the power of attack, the advantage of mage's skill is the power of defense, and the advantage of priest's skill is the ability to cure illness. An avatar can be a warrior, a priest, or a mage. To improve their skills, the avatars cooperate with friends to accomplish tasks. The more friends, the higher efficiency. Two individuals $i$ and $m$ are allowed to establish social ties to satisfy their desire of making friends and enhance their utility of collaborations. The strength of social ties is measured by the intimacy $I_{i,m}$, which increases according to the collaborative activities of $i$ and $m$ if they belong to the same society; otherwise, $I_{i,m}$ remains zero if $i$ and $m$ belong to two different societies. Hence the friendship networks of the two camps are essentially separated. As a measure of closeness to each friendship, the values of intimacy are recorded every day. When two individuals in the same society form a team and collaborate to accomplish a task, their intimacy increases. The evolving intimacy allows us to track the evolution of the cooperation behavior in the socioeconomic networks. Each individual can maintain a friendship list, denote as $\mathcal{F}_i$ for individual $i$. The social tie is symmetric: if $i \in \mathcal{F}_m$, then $m \in \mathcal{F}_i$.  In addition to the friendship network, the game data contains other socioeconomic networks, such as the face to face trading networks between initiators and receivers, the vendor trading networks between vendors and costumers, the mail networks between senders and receivers, the mentor networks between students and mentors, the kill networks between killers and victims. In this paper, our focus is the friendship network. More details of the database can be found in our earlier works about the triadic motifs in dependence networks \cite{Xie-Li-Jiang-Zhou-2014-SR} and skill complementarity in collaboration networks \cite{Xie-Li-Jiang-Tan-Podobnik-Zhou-Stanley-2016-SR}.

\subsection{Economic outputs of individuals}
%
%
We measure the economic {{output by}} converting the virtual money and items into a standardized currency for each individual.
There are two virtual currencies, {\emph{Xingbi}} and
{\emph{Jinbi}}. {The {\emph{Xingbi}} and {\emph{Jinbi}} can be exchanged in the built-in platform in each virtual world}. {{The}} virtual system has an approximately
stable exchange rate between {\emph{Xingbi}} and the Chinese currency
{\emph{Renminbi}}. {{\emph{Jinbi}} is
produced by the economic activities of the individuals when they form a team and collaborate to accomplish the tasks. The currency {\emph{Jinbi}} and virtual items, such as weapons, clothes, and medicines, are awarded to individuals when monsters are killed and tasks are accomplished.}

We convert the produced items and {\emph{Jinbi}} to
{\emph{Xingbi}} to obtain the real economic {{output}} for each individual on
each day. On average, the {{normal life span}} of virtual societies is close to 5 months \cite{Xie-Li-Jiang-Tan-Podobnik-Zhou-Stanley-2016-SR}.  Therefore, we calculate the {{output}} of each individual for a fixed period of 145 days for all virtual societies, denoted as $y_{i}=\ln\sum_{t=1}^{t=145}y_{i,t}$.

\begin{figure*}[h!]
  \centering
 \includegraphics[width=15cm]{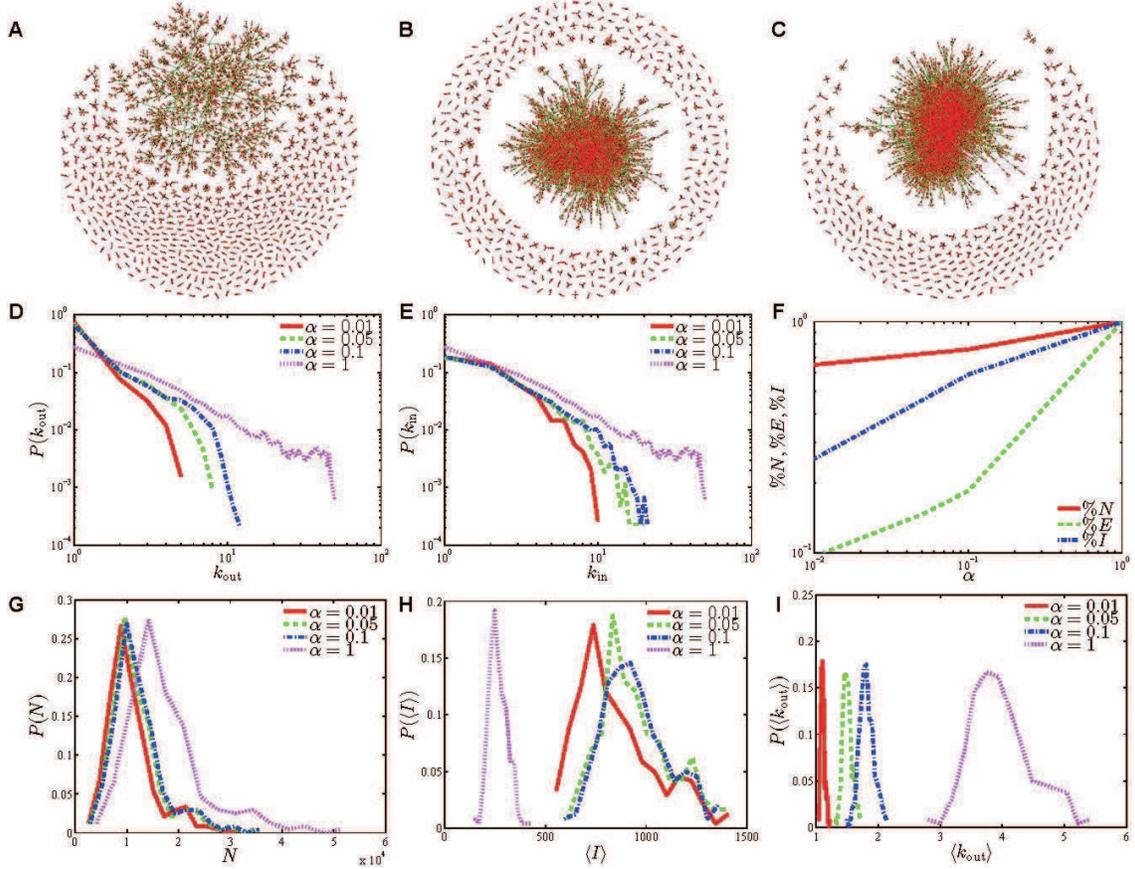}
    \caption{\label{Fig:XCB:DependNet}Illustration of dependence networks and { basic statistics of network structure}. { (A) (B) (C) Plots of dependence networks from a virtual society with significant level $\alpha=0.01, 0.05, 0.1$ respectively. (D) Out-degree distribution $P(k_{\rm{out}})$ of the dependence networks in (A) (B) (C). (E) In-degree distribution $P(k_{\rm{in}})$ of the dependence networks in (A) (B) (C). (F) Fraction of nodes, edges, weight kept in the dependence networks as a function of the significant level $\alpha$ in the filters. (G) The distribution $P(N)$ of 248 dependence networks' sizes $N$. (H) The distribution $P(\langle I\rangle)$ of the average intimacy of the friendship in the 248 dependence networks. (I) The distribution $P(\langle k_{\rm{out}}\rangle)$ of the average degree of the 248 dependence networks.}}
\end{figure*}

\subsection{Construction of dependence network}

For a given friendship network, we construct a dependence network by removing the insignificant edges based on statistical validation~\cite{Serrano-Boguna-Vespignani-2009-PNAS}. {First, we define the relative intimacy of individual $m$ in reference to all friends of individual $i$ as $w_{i,m} = I_{i,m}/\sum_{m=1}^{N}I_{i,m}$.} Obviously, $w_{i,m} \neq w_{m,i}$.
Following the statistical validation~\cite{Serrano-Boguna-Vespignani-2009-PNAS}, a directed tie $i \rightarrow m$ is significant at the level of $\alpha$ if
\begin{equation}
   \alpha_{i,m} =  1-(k_{i}-1)\int_{0}^{w_{i,m}}(1-x)^{k_{i}-2}{\rm{d}}x < \alpha,
    \label{Eq:Edges:Alpha:ij}
\end{equation}
where $k_{i}$ is the degree of individual $i$. { If $\alpha_{m,i}< \alpha$, the directed tie $m \rightarrow i$ is significant. For each society, we set the significant level $\alpha$ and remove the insignificant links,} resulting a directed dependence network. If link $i \to m $ is significant, it means that individual $m$ is relatively important to individual $i$ in $i$'s friends. { In other words, individual $i$ depends individual $m$. By the disparity filter~\cite{Serrano-Boguna-Vespignani-2009-PNAS}, the dependence networks are the backbones of the original friendship networks.}

{
Fig. \ref{Fig:XCB:DependNet} (A, B, C) show the topological structure of the dependence network constructed from a virtual society with significant level $\alpha=0.01, 0.05, 0.1$ respectively.
We analysis the out-degree distribution $P(k_{\rm{out}})$ and in-degree distribution $P(k_{\rm{in}})$ of dependence networks for different significant level $\alpha=0.01, 0.05, 0.1$ in Fig. \ref{Fig:XCB:DependNet} (D, E).
The degree distribution of the dependence network is obviously different from the original friendship network ($\alpha=1$). Using the smaller significance level $\alpha$, the disparity filter reduces more edges. For different significant level $\alpha$, the fraction of nodes, edges, weight kept in the dependence network increase with the significant level $\alpha$ as shown in Fig. \ref{Fig:XCB:DependNet} (F).
And the average degree of the 248 dependence networks is monotone increasing with the significant level $\alpha$ in Fig. \ref{Fig:XCB:DependNet} (I).
 For different significant level $\alpha=0.01, 0.05, 0.1$, the distribution $P(N)$ of the 248 dependence networks' sizes $N$ have nearly the same shape in Fig. \ref{Fig:XCB:DependNet} (G). The numbers of the avatars range from thousand to fifty thousands.
Comparing with the global threshold filter, the disparity filter considers the relevant edges and ensures that the edges with small intimacy are not neglected.
So we can find that the average intimacy kept in the dependence network is not monotone decreasing with the significant lever $\alpha$ in the Fig. \ref{Fig:XCB:DependNet} (H).
}

\subsection{Quantifying position ratio profile}

\begin{figure}[h!]
\centering
\includegraphics[width=10cm]{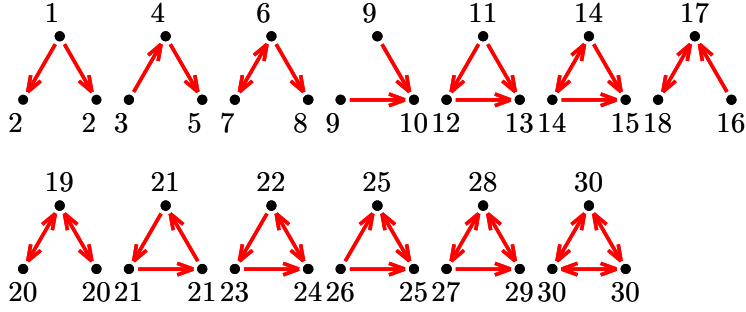}
\caption{Plots of 30 unique positions $j\in \{1,2,...,30\}$ in 13 directed triadic motifs.}
\label{Fig:XCB_Motifs_Position30}
\end{figure}

Following the method~\cite{Milo-ShenOrr-Itzkovitz-Kashtan-Chklovskii-Alon-2002-Science} to identify motifs, we are able to identify 13 different directed triadic motifs, as shown in Fig.~\ref{Fig:XCB_Motifs_Position30}, in dependence networks.
These motifs uncover {{the}} dependence structures among individuals at the microscopic level. For example, motif \textcircled{\scriptsize{2}}$\leftarrow$\textcircled{\scriptsize{1}}$\rightarrow$\textcircled{\scriptsize{2}} stands for the situation that one individual depends on the other two individuals, motif
 \textcircled{\scriptsize{3}}$\rightarrow$\textcircled{\scriptsize{4}}$\rightarrow$\textcircled{\scriptsize{5}}
 means that one individual depends on another individual which in turn depends on the third individual, {{whereas}} motif
  \textcircled{\scriptsize{a}}$\rightarrow$\textcircled{\scriptsize{b}}$\rightarrow$\textcircled{\scriptsize{c}}$\rightarrow$\textcircled{\scriptsize{a}}
  represents the situation that individual $a$ depends on individual $b$, individual $b$ depends on individual $c$, and individual $c$ also depends on individual $a$.
We can further locate 30 distinct motif positions \cite{Stouffer-SalesPardo-Sirer-Bascompte-2012-Science} within the 13 different directed triadic motifs, as shown in Fig.~\ref{Fig:XCB_Motifs_Position30}. We directly enumerate the relative frequency $p_{i,j}$ that individual $i$ appears in
position $j$ across the 13 motifs, which gives the motif position ratio profile $p_i=(p_{i,1},p_{i,2},...,p_{i,30})$ for individual $i$. Note that we have $\sum_{j=1}^{30}p_{i,j}=1$.

Hence, we first define the $z$-sore of occurrence frequency for position $j$:
\begin{equation}
Z_{i,j}=\frac{p_{i,j}-\langle p_{i,j}\rangle}{\sigma(p_{i,j}) },
    \label{Eq:Positions_Zscore}
\end{equation}
where $\langle p_{i,j}\rangle=\sum_{i=1}^{N}p_{i,j}/N$ and $\sigma(p_{i,j})$
are the mean and standard deviation of $p_{i,j}$ over all individuals in the dependence network of a given virtual society \cite{Milo-Itzkovitz-Kashtan-Levitt-ShenOrr-Ayzenshtat-Sheffer-Alon-2004-Science, Milo-Kashtan-Itzkovitz-Newman-Alon-2004-XXX}.
The structural similarity between individual $i$ and $m$ are thus defined as the correlation coefficients between $Z_i$ and $Z_m$, such that,
\begin{equation}
s_{i,m}=\frac{E[(Z_{i,j}-\langle Z_{i}\rangle)(Z_{m,j}-\langle Z_{m}\rangle)]}{E[(Z_{i,j}-\langle Z_{i}\rangle)^2]^{1/2}E[(Z_{m,j}-\langle Z_{m}\rangle)^2]^{1/2}},
    \label{Eq:Positions_sij}
\end{equation}
where $E[x]$ is the mathematical expectation of $x$.

\subsection{Classifying individuals based on their position ratio profile}

We employ the $k$-means algorithm to classify $N$ individuals based on their position ratio {{profiles $\{Z_i\}$}} in the dependence social networks.
The position ratio profile of individual $i$ is wrote as $Z_i=(Z_{i,1},Z_{i,2},...,Z_{i,30})$.
For each virtual society, we can get a matrix of position ratio profile, denoted by $\cal{Z}$, in which
rows correspond to individuals and columns correspond to positions in triadic motifs.
By adopting {{the}} $k$-means algorithm, we partition the individuals into $k$ clusters in terms of minimizing the sum, over all clusters, of the within-cluster sums of point-to-cluster-centroid distances. The partition algorithm is implemented in an iterative way.
{
To evaluate the optimal number of clusters, we calculate the clustering evaluation object containing Davies-Bouldin index values
and find the minimum criterion value. Then we can get the optimal number of clusters $k$.}
The correlation distance defined as one minus the sample correlation between points (treated as $Z_i$) is used to capture the closeness between individuals in $k$-means algorithm, such as,
\begin{equation}
d_{i,m}=1-s_{i,m}.
    \label{Eq:Distance_sij}
\end{equation}
{{The}} $k$-means algorithm returns the cluster indices of each individual and the $k$ cluster centroid locations in a $k$-by-$30$ matrix.
Each centroid is the component-wise mean of the points in that cluster.
We denote the $k$-th cluster centroid locations as $P_{k} =(P_{k,1},P_{k,2},...,P_{k,30})$, where $P_{k,j}=\sum_{i\in {\cal{C}}_{k}}p_{i,j}/c_k$ and $j\in \{1,2,...,30\}$. $c_k$ is the number of individuals in cluster ${\cal{C}}_k$. We have $\sum_{j=1}^{30}P_{k,j}=1$, and $P_{k}$ is the position ratio profile of cluster $k$.

\section{Results}

\subsection{Quantifying position ratio profile}

Individuals inside the virtual societies are involved in all kinds of social and economic activities. The theory of {structural} holes already tells us that the agent occupies special position in social network will have great influence on his economic performance \cite{Burt-2009}. Here, we empirically investigate the underlying connections between network structure and economic {{output}} for individuals in 248 virtual societies. Our data are game logs from a popular MMORPG in China (see data description in section \ref{section:Materials and Methods}). For each society, we construct a dependence network of individuals, which only keeps the multiscale backbone of original friendship network (see construction of dependence network in section \ref{section:Materials and Methods}). Fig.~\ref{Fig:XCB:DependNet} illustrates the dependence network from a virtual society { and some basic
 statistics about the topological structure.} One can see that there are lots of in-stars in dependence network, which is in accordance with the findings \cite{Xie-Li-Jiang-Zhou-2014-SR} that the in-degree distribution is much fatter than out-degree distribution.
Such phenomena can be explained by that in virtual societies the individuals with high levels may play a relatively important role in friend lists of low level individuals, which is supported by the strong preference of connecting high level individuals for low level individuals \cite{Xie-Li-Jiang-Zhou-2014-SR}.

\begin{figure*}[h!]
  \centering
 \includegraphics[width=12cm]{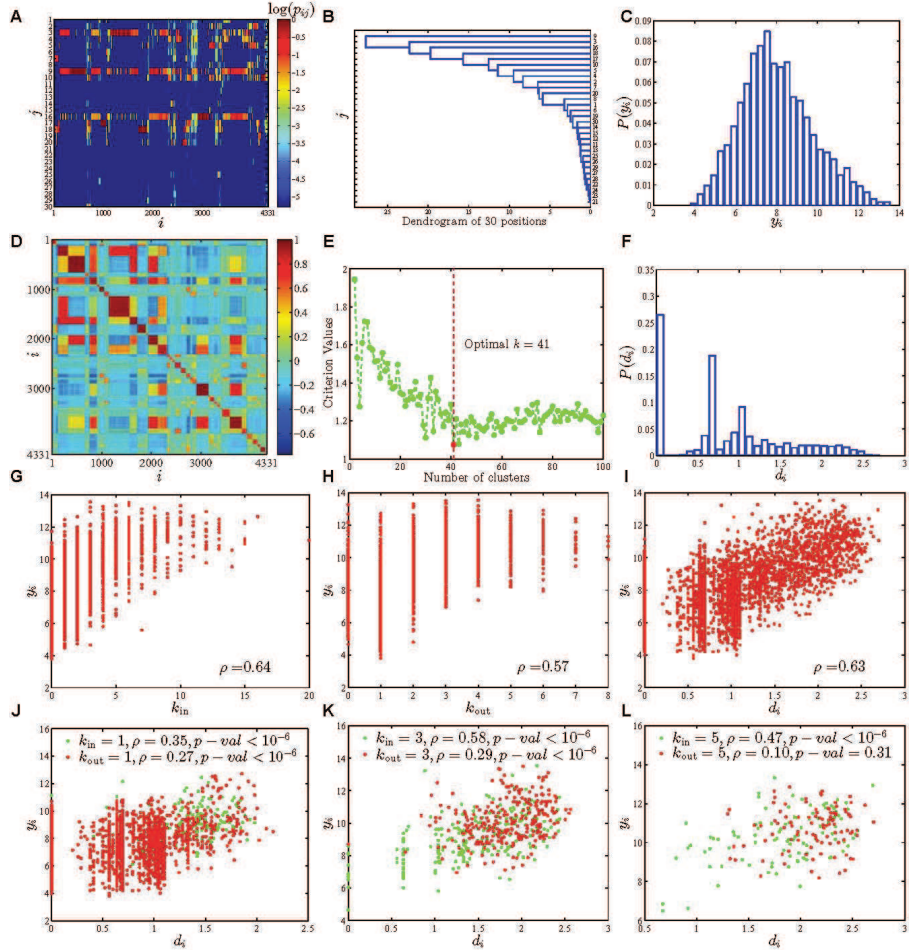}
  \caption{\label{Fig:XCB_Motifs_M30Position_Block} Results from the analysis of position ratio profiles for individuals in Fig.~\ref{Fig:XCB:DependNet} { (B) with significant level $\alpha=0.05$.}
(A) Plots of position ratio profiles for all individuals. The color bar represents the {log} value of position ratio profiles. {
(B) Dendrogram of the 30 position based on the position ratio profiles for individuals in Fig.~\ref{Fig:XCB:DependNet} (B).
(C) Distribution of the economic {{output}} $y_i$ of individuals in Fig.~\ref{Fig:XCB:DependNet} (B).
(D) Plots of the similarities $s_{i,m}$ for all individuals. Similarity is defined as the correlation of position ratio profiles between individual $i$ and $m$. The color bar stands for the value of similarities $s_{i,m}$.
(E) Evaluate the optimal number of clusters $k$ by using the Davies-Bouldin criterion.
(F) Distribution of the individual position diversity $d_i$ in Fig.~\ref{Fig:XCB:DependNet} (B).
(G, H, I) Plots of the relations between the economic {{output}} and the in-degree $k_{\rm{in}}$, out-degree $k_{\rm{out}}$, position diversity $d_i$.
(J, K, L) Plots of the relations between the economic {{output}} and the position diversity $d_i$ of individuals with $k_{\rm{in}}, k_{\rm{out}}= 1, 3, 5$.
 }}
\end{figure*}

For a given dependence network, we can estimate a position ratio profile for each individual (see quantifying position ratio profile in section \ref{section:Materials and Methods}).
Xie et al. revealed that the open motifs have higher occurrence frequency than close motifs \cite{Xie-Li-Jiang-Zhou-2014-SR}, suggesting that the positions in the open motifs may have higher occurrence frequency than positions in close motifs. Fig.~\ref{Fig:XCB_Motifs_M30Position_Block} (A) shows the position ratio profile for all individuals in dependence networks in Fig.~\ref{Fig:XCB:DependNet}.
{ There are more zeros in the position ratio profiles of individuals with low degrees than those with high degrees.}
More extremely, there are lots of individuals occupying only one specific position ($j$ = 3, 9, 10, 16, 17, and 18) of triadic motifs in dependence networks, which leads to $p_{i, j} = 1$. The individuals with lager value of $p_{i,3},p_{i,9},p_{i,16}$ in their position ratio profile may correspond to leaf nodes in dependence networks.
Note that the occurrence frequency of motif \textcircled{\scriptsize{a}}$\rightarrow$\textcircled{\scriptsize{b}}$\rightarrow$\textcircled{\scriptsize{c}}$\rightarrow$\textcircled{\scriptsize{a}}
 is 0, resulting in that the position 21 is impossible to be observed.

\subsection{Similarity of position ratio profiles between individuals}

The triadic motifs are regarded as building blocks of complex networks \cite{Milo-ShenOrr-Itzkovitz-Kashtan-Chklovskii-Alon-2002-Science}, suggesting that the 30 positions in triadic motifs may contain important {{structural}} information {{of}} nodes in complex {{networks}}.
Here we utilize the occurrence frequency of the 30 different positions in dependence networks to represent the network structure profile for each individual. Based on these profiles, we can assess the structural similarity between individuals, which allows us further to group individuals.
In dependence networks, it is observed that some motifs, for example,  \textcircled{\scriptsize{3}}$\rightarrow$\textcircled{\scriptsize{4}}$\rightarrow$\textcircled{\scriptsize{5}}, \textcircled{\scriptsize{9}}$\rightarrow$\textcircled{\scriptsize{10}}$\leftarrow$\textcircled{\scriptsize{9}}, \textcircled{\scriptsize{16}}$\rightarrow$\textcircled{\scriptsize{17}}$\rightleftharpoons$\textcircled{\scriptsize{18}}, appears more frequency than other motifs \cite{Xie-Li-Jiang-Zhou-2014-SR}.

By ordering the individuals according to the rule that nodes with large similarity are close to each other, we plot the structural similarity in Fig.~\ref{Fig:XCB_Motifs_M30Position_Block} (D). The color bar stands for the value of similarities $s_{i,m}$. One intriguing observation is that there is a block-diagonal structure, strongly indicating the function of grouping individuals for position ratio profiles. This inspires us to further classify the individuals into clusters by $k$-means algorithm (see classifying individuals based on their position ratio profile in section \ref{section:Materials and Methods}).  { In Fig.~\ref{Fig:XCB_Motifs_M30Position_Block}(A) and Fig.~\ref{Fig:XCB_Motifs_M30Position_Block}(D),
there are some classes which are made up of some special individuals, such as individuals with $p_{i,3}=1$
in motif \textcircled{\scriptsize{3}}$\rightarrow$\textcircled{\scriptsize{4}}$\rightarrow$\textcircled{\scriptsize{5}},
or individuals with  $p_{i,9}=1$ or $p_{i,10}=1$
in motif \textcircled{\scriptsize{9}}$\rightarrow$\textcircled{\scriptsize{10}}$\leftarrow$\textcircled{\scriptsize{9}},
or individuals with $p_{i,16}=1$, $p_{i,17}=1$, or $p_{i,18}=1$
in motif \textcircled{\scriptsize{16}}$\rightarrow$\textcircled{\scriptsize{17}}$\rightleftharpoons$\textcircled{\scriptsize{18}}.
}

The $k$-means algorithm is an iterative partitioning algorithm, which maximizes the similarity of the within-cluster.
{ To evaluate the optimal number of clusters, we create a clustering evaluation object containing  Davies-Bouldin index values in Fig.~\ref{Fig:XCB_Motifs_M30Position_Block} (E).
From the minimum criterion value, we can get the optimal number of clusters $k=41$.
Based on the definition of position ratio profiles, there is a correlation among the 30 positions. From the dendrogram of the 30 position in Fig.~\ref{Fig:XCB_Motifs_M30Position_Block} (B), we can find that a avatar at position 22 has a higher chance to be also in position 23 or 24 in the same motif. But there is no obvious classification among the 30 position.  The block-diagonal structure can not be explained by this relation among the 30 position.  An alternative method to elucidate the position significant profiles is to calculate the difference of position ratio profiles between the dependence networks and the reference
randomized dependence networks \cite{Milo-ShenOrr-Itzkovitz-Kashtan-Chklovskii-Alon-2002-Science}. But this method is too computationally intensive for our 248 dependence networks and some of the networks' sizes
are larger than fifty thousands.
}

\subsection{Relationship between position ratio profiles and economic outputs}

In social network, the individuals in key positions may have great impacts of network dynamics if some controlling strategies are applied on these key nodes.
This leads to the conjecture that the position ratio profiles may have potential influence on the economic outputs for individuals.  We estimate the economic {{output}} $y_{i}$ for each individual in  a given world (see economic outputs of individuals in section \ref{section:Materials and Methods}).
To make these economic outputs comparable between different virtual worlds, we standardize the economic {{output}} within each society.
The standardization of economic performance of individual $i$ is defined as {{output}} per capita,
$Y_{i}=[y_{i}-\langle y_{i}\rangle]/\sigma(y_{i})$,
where $\langle y_{i}\rangle$ and $\sigma(y_{i})$
are the mean and standard deviation of $y_{i}$ over all individuals in a given society.
Therefore, the mean of economic performance of the class ${\cal{C}}_k$ can be estimated via $Y_{k}=\sum_{i\in {\cal{C}}_k}Y_{i}/c_{k,}$, where $c_{k}$ is the number of individuals in class ${\cal{C}}_k$.

Motivated by social diversity~\cite{Eagle-Macy-Claxton-2010-Science}, we define the individual position diversity $d_{i}$ as a function of Shannon entropy to quantify how individual appears in the 30 positions:
\begin{equation}
 d_{i}=-\sum_{j=1}^{30}p_{i,j}\ln p_{i,j}.
    \label{Eq:Positions_H}
\end{equation}

\begin{figure*}[h!]
\centering
\includegraphics[width=15cm]{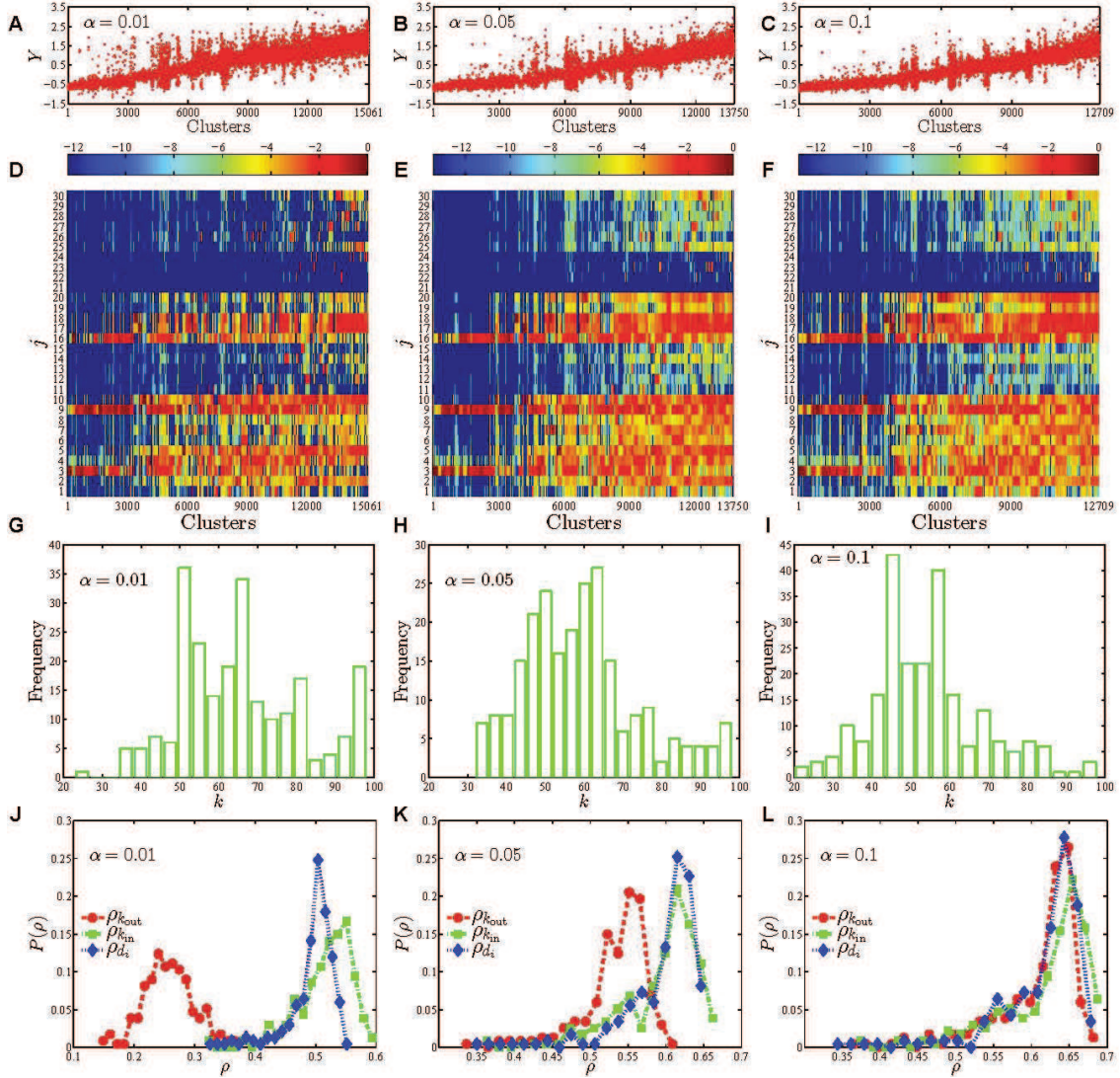}
\caption{\label{Fig:XCB_Motifs_M30Position_Block_Ys}Relation between position diversity and economic {{output}}
for clusters in the dependence network of 248 virtual societies.  {
  The three columns from left to right correspond to the dependence networks from all the virtual society with significant level $\alpha=0.01, 0.05, 0.1$ respectively. The four rows from top to bottom correspond to (A, B, C) the plots of the average economic {{output}} of clusters,
(D, E, F) the plots of the value of cluster position ratio profiles $P_k$, (G, H, I) the distribution of the optimal number of clusters $k$, (J, K, L) the distribution of correlation coefficients between the economic {{output}} and the individuals' in-degree $k_{\rm{in}}$, out-degree $k_{\rm{out}}$, position diversity $d_i$. }}
\end{figure*}

{  Fig.~\ref{Fig:XCB_Motifs_M30Position_Block} shows the results from the analysis of position ratio profiles for individuals in Fig.~\ref{Fig:XCB:DependNet} with significant level $\alpha=0.05$.
The position ratio profiles of 4331 individuals are presented in Fig.~\ref{Fig:XCB_Motifs_M30Position_Block} (A). To improve the visualization, the color bar represents the log value of position ratio profiles $p_{i,j}$.
The position diversity $d_{i}=0$ means that the individual only occupy one triadic motif position.
As shown in Fig.~\ref{Fig:XCB_Motifs_M30Position_Block} (A), there are individuals with $p_{i,3}=1$ in motif \textcircled{\scriptsize{3}}$\rightarrow$\textcircled{\scriptsize{4}}$\rightarrow$\textcircled{\scriptsize{5}},
$p_{i,9}=1$, $p_{i,10}=1$ in motif \textcircled{\scriptsize{9}}$\rightarrow$\textcircled{\scriptsize{10}}$\leftarrow$\textcircled{\scriptsize{9}},
$p_{i,16}=1$, $p_{i,17}=1$, and $p_{i,18}=1$ in motif \textcircled{\scriptsize{16}}$\rightarrow$\textcircled{\scriptsize{17}}$\rightleftharpoons$\textcircled{\scriptsize{18}}.
Fig.~\ref{Fig:XCB_Motifs_M30Position_Block} (C, F) show the economic {{output}} $y_i$ and the individual position diversity $d_i$ respectively.
The individuals with position diversity $d_{i}=0$ account for a large part of avatars in the dependence network of a given society.
Fig.~\ref{Fig:XCB_Motifs_M30Position_Block} (L) shows the scatter plot of economic {{output}} $y_{i}$ and position diversity $d_{i}$.
The correlation coefficient $\rho$ is 0.63 and $p$-value is less than $10^{-6}$, implying that the correlation between $y_{i}$ and $d_{i}$ is positive and highly significant. Our results indicate that the individuals who appear in more triadic motif positions have higher economic outputs.
It is easy to find that more active players have more friends (more in-degree and out-degree) and appear in more triadic motif positions.
Here we show the plots of the relation between the economic {{output}} $y_i$ and the individuals' in-degree $k_{\rm{in}}$, out-degree $k_{\rm{out}}$ in Fig.~\ref{Fig:XCB_Motifs_M30Position_Block} (G, H) respectively. This agrees with Fuchs' result \cite{Fuchs-Thurner-2014-PLoS1}
 that the output is correlated to both in- and out-degree.
 To avoid the impact of the in- and out-degree, we investigate the
 relation between the economic {{output}} $y_i$ and the position diversity $d_i$ of individuals with fixed $k_{\rm{in}}, k_{\rm{out}}= 1, 3, 5$ in
 Fig.~\ref{Fig:XCB_Motifs_M30Position_Block} (J, K, L).
And we can get the same conclusion that the individual position diversity increases economic {{output}} of individuals with fixed $k_{\rm{in}}$ or $k_{\rm{out}}$.
 What is more, we also find that the economic {{output}} is susceptible to asymmetries between individuals' in- and out-degree in our dependence socioeconomic network. The individuals with high in-degree have higher economic outputs than with high out-degree. Because it is difficult to get the records and measure the activity of individuals accurately, we assume that the individuals with the same in- and out-degree have almost the same activity.
 }

The $k$-means algorithm also gives cluster centroid locations denoted as $P_{k} =(P_{k,1},P_{k,2},...,P_{k,30})$ for each class ${\cal{C}}_{k}$. Similar to Eq.~\ref{Eq:Positions_H}, we define the cluster position diversity as $D_{k}=-\sum_{j=1}^{30}P_{k,j}\ln P_{k,j}$.
{  In Fig.~\ref{Fig:XCB_Motifs_M30Position_Block_Ys}(A-F), the clusters are sorted by the cluster position diversity $D_{k}$ in ascending order.
Considering the impact of different significant level $\alpha$ on the relation between the economic {{output}} and position diversity, we analysis
the dependence network from all the virtual society with different significant level $\alpha=0.01, 0.05, 0.1$.
 In Fig.~\ref{Fig:XCB_Motifs_M30Position_Block_Ys}, the three columns from left to right correspond to the dependence networks  with significant level $\alpha=0.01, 0.05, 0.1$ respectively. Fig.~\ref{Fig:XCB_Motifs_M30Position_Block_Ys} (A, B, C) show the plots of the average economic {{output}} $Y$ of classes and Fig.~\ref{Fig:XCB_Motifs_M30Position_Block_Ys} (D, E, F) illustrate the plots of the value of $P_k$ for different significant level $\alpha=0.01, 0.05, 0.1$ respectively. The clusters are sorted by the cluster position diversity $D_{k}$ in ascending order, so one can see that the economic {{output}} $Y$ increases with the cluster position diversity $D_{k}$.  We can find that lots of clusters are comprised of individuals with $P_{k,3},P_{k,9},P_{k,16}$ equalling to one.
Although these individuals have the same individual position diversity $d_i=0$, their economic outputs could be different. For each dependence network with a given $\alpha$, we calculate the Davies-Bouldin index values to evaluate the optimal number of clusters and get the optimal number of clusters $k$.
 The distributions of the optimal number $k$ are shown in Fig.~\ref{Fig:XCB_Motifs_M30Position_Block_Ys} (G, H, I). The
 average optimal number $k$ is between 50 and 60.
 }

 { To compare the relation coefficients between the economic {{output}} and the individuals' in-degree $k_{\rm{in}}$, out-degree $k_{\rm{out}}$, position diversity $d_i$, we denote the coefficients as $\rho_{k_{\rm{in}}}$, $\rho_{k_{\rm{out}}}$ and $\rho_{d_{i}}$ respectively and  draw the distribution of the three kinds of coefficients for different significant level $\alpha$.
When the significant levels $\alpha$ is smaller, there are greater difference between the three kinds of coefficients.
At the same time, the asymmetries between individuals' in- and out-degree in our dependence socioeconomic network is obvious with the significant level $\alpha=0.01$.}

\subsection{Economic {{output}} of individuals with the position diversity $d_{i}=0$}

\begin{figure*}[h!]
\centering
\includegraphics[width=12cm]{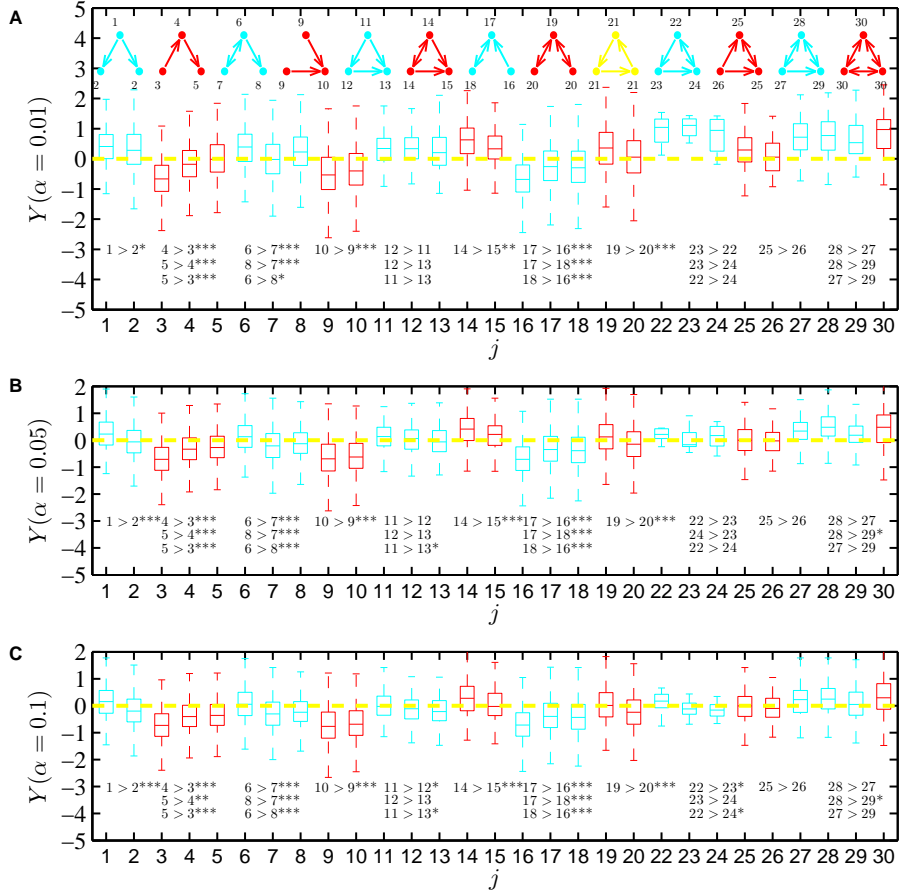}
\caption{\label{Fig:XCB_Motifs_M30Position_Diversity_Ys_H}Different economic {{output}} of individuals with the same position diversity $d_{i}=0$ in the dependence networks from all the virtual societies with significant level $\alpha=0.01, 0.05, 0.1$.
{
Boxplots of economic {{output}} of individuals with position diversity $d_{i}=0$ for 30 unique positions $j\in \{1,2,...,30\}$ in 13 directed triadic motifs.
The colors are only visually easier to distinguish the positions in the two adjacent motifs. There is no practical meaning.
We perform $t$-tests on pairs of economic {{output}} of the individuals at two position in the same motif.
The mean economic {{output}} of individual at position $1$ is greater than that of individuals within position $2$ in motif \textcircled{\scriptsize{2}}$\leftarrow$\textcircled{\scriptsize{1}}$\rightarrow$\textcircled{\scriptsize{2}}.
This relationship can be expressed as $1>2$.
The superscript *, **, *** stand that the relationship is acceptable at the significant level $0.05$, $0.001$, $0.0001$ respectively.} }
\end{figure*}

Here, we conduct a comparison of the economic {{output}} of individuals with individual position diversity $d_{i}=0$ in the dependence networks from all the virtual society with significant level $\alpha=0.01, 0.05, 0.1$.
The individual position diversity $d_{i}=0$ means that the individual appears only in one triadic motif position.
 {
We perform $t$-tests on pairs of economic {{output}} of the individuals at two position in the same motif.}
Fig.~\ref{Fig:XCB_Motifs_M30Position_Diversity_Ys_H} shows the box plots of economic {{output}} of individuals with position diversity $d_{i}=0$ for 30 unique positions $j\in \{1,2,...,30\}$ in 13 directed triadic motifs.
 {
In Fig.~\ref{Fig:XCB_Motifs_M30Position_Diversity_Ys_H}, the superscript *, **, *** stand that the relationship is acceptable at the significant level $0.05$, $0.001$ and $0.0001$ respectively. For all the dependence networks with significant level $\alpha=0.01, 0.05, 0.1$, the results of the $t$-tests are similar. }

 {In Fig.~\ref{Fig:XCB_Motifs_M30Position_Diversity_Ys_H},} we perform $t$-tests on pairs of individual economic {{output}} of the individuals at two position { in the dependence networks.}
It is worthy stressing that the mean economic {{output}} of individual at position $1$ is greater than that of individuals within position $2$ in motif \textcircled{\scriptsize{2}}$\leftarrow$\textcircled{\scriptsize{1}}$\rightarrow$\textcircled{\scriptsize{2}}.
 {This relationship can be expressed as $1>2$ in Fig.~\ref{Fig:XCB_Motifs_M30Position_Diversity_Ys_H}.}
The mean economic {{output}} of individuals at position $4$  and $5$ are greater than that of individuals at position $3$ in motif
\textcircled{\scriptsize{3}}$\rightarrow$\textcircled{\scriptsize{4}}$\rightarrow$\textcircled{\scriptsize{5}}.
 {These relationship can be expressed as $4>3$ and $5>3$ in Fig.~\ref{Fig:XCB_Motifs_M30Position_Diversity_Ys_H}.}
The mean economic {{output}} of individuals at position $10$ is greater than that of individuals at position $9$ in motif
\textcircled{\scriptsize{9}}$\rightarrow$\textcircled{\scriptsize{10}}$\leftarrow$\textcircled{\scriptsize{9}}.
 {This relationship can be expressed as $10>9$ in Fig.~\ref{Fig:XCB_Motifs_M30Position_Diversity_Ys_H}.}
The mean economic {{output}} of individuals at position $17$  and $18$ are greater than that of individuals at position $16$ in motif
\textcircled{\scriptsize{16}}$\rightarrow$\textcircled{\scriptsize{17}}$\rightleftharpoons$\textcircled{\scriptsize{18}}.
 {These relationship can be expressed as $17>16$ and $18>16$ in Fig.~\ref{Fig:XCB_Motifs_M30Position_Diversity_Ys_H}.}

{{Comparing the output}} for individuals at different positions { in the dependence networks with significant level $\alpha=0.05$}, the 30 positions can be classified into two groups. The first group contains the
positions:
\textcircled{\scriptsize{3}}, \textcircled{\scriptsize{16}}, \textcircled{\scriptsize{9}}, \textcircled{\scriptsize{10}}, \textcircled{\scriptsize{18}}, \textcircled{\scriptsize{4}}, \textcircled{\scriptsize{17}}, \textcircled{\scriptsize{5}}, \textcircled{\scriptsize{7}}, \textcircled{\scriptsize{20}}, \textcircled{\scriptsize{8}}, \textcircled{\scriptsize{13}}, \textcircled{\scriptsize{2}}, \textcircled{\scriptsize{26}}
.
 The second group contains positions:
\textcircled{\scriptsize{25}}, \textcircled{\scriptsize{23}}, \textcircled{\scriptsize{12}}, \textcircled{\scriptsize{24}}, \textcircled{\scriptsize{11}}, \textcircled{\scriptsize{19}}, \textcircled{\scriptsize{6}}, \textcircled{\scriptsize{15}}, \textcircled{\scriptsize{29}}, \textcircled{\scriptsize{22}}, \textcircled{\scriptsize{1}}, \textcircled{\scriptsize{27}}, \textcircled{\scriptsize{30}}, \textcircled{\scriptsize{14}}, \textcircled{\scriptsize{28}}. All the positions are sorted by means of individual economic {{output}} in ascending order. The mean economic {{output}} of individuals in the first group is less than 0, while in the second group the mean {{output}} is greater than 0.
Such differences could result from that most of positions of the first group are in open motifs while most of positions in second group belong to close motifs.

\section{Discussion}

In this paper we analyzed the relationship between friendship structure and economic {{output}} for individuals in a massively multiplayer online role-playing game. We {{found}} that the individual position diversity of individuals is positively correlated with their economic output and social status.  We have developed a new approach to study structural similarity of individuals in networks and classify individuals into different clusters based on their position ratio profiles. It is found that the economic {{output}} of leaf nodes {{is}} significantly lower than the nodes at the other triadic motif positions. The individual position diversity positively correlates with the economic {{output}}. Our results are in consistent with the results from the physical society \cite{Bhattacharya-Dugar-2014-MS,Ball-Newman-2013-NS}. Furthermore, Eagle et al. found that the diversity of individual relationships within a community strongly correlates with the economic development of the community \cite{Eagle-Macy-Claxton-2010-Science}. In a recent analysis, Xie et al. found the degree of skill complementarity is positively correlated with their output \cite{Xie-Li-Jiang-Tan-Podobnik-Zhou-Stanley-2016-SR}.

Much evidence shows that the behaviour of individuals in virtual society {{is}} representative in many aspects of human behaviour in physical society. This is rational because individuals' decision process is determined by players that are real human being. The invisible hand operates not only in modern societies but also in ancient societies, not only in real societies but also in virtual societies.
Yee et al. argued that online environments such as MMORPGs could potentially be unique research platforms for the social sciences and clinical therapy, but it is crucial to firstly establish that social behavior and norms in virtual environments are comparable to those in the physical world \cite{Yee-Bailenson-Urbanek-Chang-Merget-2007-CPB}.
To investigate the relation between friendship (or socioeconomic) networks in virtual and real worlds, Grabowski and Kruszewska conducted a survey among the players of an online game and construct the off-line network \cite{Grabowski-Kruszewska-2007-IJMPC}. They showed that the structure of the friendship network in virtual world was very similar to the structure of different social networks in real world.

We believe that our interdisciplinary work represents significant scientific evidence for understanding the behaviour of social systems. It involves topics that range from social science, network science, and economics to human dynamics. It also enriches our understanding on the formation of socio-economic networks and proposes a new method to classify nodes of complex networks for understanding of people's economic behaviour from the big data of massive players. Our work sheds new light on the scientific research utility of virtual worlds for studying human behaviour in complex socio-economic systems.

\begin{acknowledgments}
  This work was supported by the National Natural Science Foundation of China [11505063, 11375064 and 11605062];
the Shanghai Chenguang Program [15CG29]; the Ph.D. Programs Foundation of Ministry of Education of China [20120074120028]; and the Fundamental Research Funds for the Central Universities.
\end{acknowledgments}

\bibliography{E:/Papers/Auxiliary/Bibliography}

\begin{thebibliography}{45}
\expandafter\ifx\csname natexlab\endcsname\relax\def\natexlab#1{#1}\fi
\expandafter\ifx\csname bibnamefont\endcsname\relax
  \def\bibnamefont#1{#1}\fi
\expandafter\ifx\csname bibfnamefont\endcsname\relax
  \def\bibfnamefont#1{#1}\fi
\expandafter\ifx\csname citenamefont\endcsname\relax
  \def\citenamefont#1{#1}\fi
\expandafter\ifx\csname url\endcsname\relax
  \def\url#1{\texttt{#1}}\fi
\expandafter\ifx\csname urlprefix\endcsname\relax\def\urlprefix{URL }\fi
\providecommand{\bibinfo}[2]{#2}
\providecommand{\eprint}[2][]{\url{#2}}

\bibitem[{\citenamefont{Wang et~al.}(2009)\citenamefont{Wang, Gonz{\'a}lez,
  Hidalgo, and Barab{\'a}si}}]{Wang-Gonzalez-Hidalgo-Barabasi-2009-Science}
\bibinfo{author}{\bibfnamefont{P.}~\bibnamefont{Wang}},
  \bibinfo{author}{\bibfnamefont{M.~C.} \bibnamefont{Gonz{\'a}lez}},
  \bibinfo{author}{\bibfnamefont{C.~A.} \bibnamefont{Hidalgo}},
  \bibnamefont{and} \bibinfo{author}{\bibfnamefont{A.-L.}
  \bibnamefont{Barab{\'a}si}}, \bibinfo{journal}{Science}
  \textbf{\bibinfo{volume}{324}}, \bibinfo{pages}{1071} (\bibinfo{year}{2009}).

\bibitem[{\citenamefont{Newman}(2001)}]{Newman-2001-PNAS}
\bibinfo{author}{\bibfnamefont{M.~E.~J.} \bibnamefont{Newman}},
  \bibinfo{journal}{Proc. Natl. Acad. Sci. U.S.A.}
  \textbf{\bibinfo{volume}{98}}, \bibinfo{pages}{404} (\bibinfo{year}{2001}).

\bibitem[{\citenamefont{Bhattacharya and
  Dugar}(2014)}]{Bhattacharya-Dugar-2014-MS}
\bibinfo{author}{\bibfnamefont{H.}~\bibnamefont{Bhattacharya}}
  \bibnamefont{and} \bibinfo{author}{\bibfnamefont{S.}~\bibnamefont{Dugar}},
  \bibinfo{journal}{Manag. Sci.} \textbf{\bibinfo{volume}{60}},
  \bibinfo{pages}{1130} (\bibinfo{year}{2014}).

\bibitem[{\citenamefont{Xie et~al.}(2014)\citenamefont{Xie, Li, Jiang, and
  Zhou}}]{Xie-Li-Jiang-Zhou-2014-SR}
\bibinfo{author}{\bibfnamefont{W.-J.} \bibnamefont{Xie}},
  \bibinfo{author}{\bibfnamefont{M.-X.} \bibnamefont{Li}},
  \bibinfo{author}{\bibfnamefont{Z.-Q.} \bibnamefont{Jiang}}, \bibnamefont{and}
  \bibinfo{author}{\bibfnamefont{W.-X.} \bibnamefont{Zhou}},
  \bibinfo{journal}{Sci. Rep.} \textbf{\bibinfo{volume}{4}},
  \bibinfo{pages}{5244} (\bibinfo{year}{2014}).

\bibitem[{\citenamefont{Eagle et~al.}(2010)\citenamefont{Eagle, Macy, and
  Claxton}}]{Eagle-Macy-Claxton-2010-Science}
\bibinfo{author}{\bibfnamefont{N.}~\bibnamefont{Eagle}},
  \bibinfo{author}{\bibfnamefont{M.}~\bibnamefont{Macy}}, \bibnamefont{and}
  \bibinfo{author}{\bibfnamefont{R.}~\bibnamefont{Claxton}},
  \bibinfo{journal}{Science} \textbf{\bibinfo{volume}{328}},
  \bibinfo{pages}{1029} (\bibinfo{year}{2010}).

\bibitem[{\citenamefont{Bettencourt et~al.}(2014)\citenamefont{Bettencourt,
  Samaniego, and Youn}}]{Bettencourt-Samaniego-Youn-2014-SR}
\bibinfo{author}{\bibfnamefont{L.~M.~A.} \bibnamefont{Bettencourt}},
  \bibinfo{author}{\bibfnamefont{H.}~\bibnamefont{Samaniego}},
  \bibnamefont{and} \bibinfo{author}{\bibfnamefont{H.}~\bibnamefont{Youn}},
  \bibinfo{journal}{Sci. Rep.} \textbf{\bibinfo{volume}{4}},
  \bibinfo{pages}{5393} (\bibinfo{year}{2014}).

\bibitem[{\citenamefont{Ortman et~al.}(2015)\citenamefont{Ortman, Cabaniss,
  Sturm, and Bettencourt}}]{Ortman-Cabaniss-Sturm-Bettencourt-2015-SciAdv}
\bibinfo{author}{\bibfnamefont{S.~G.} \bibnamefont{Ortman}},
  \bibinfo{author}{\bibfnamefont{A.~H.~F.} \bibnamefont{Cabaniss}},
  \bibinfo{author}{\bibfnamefont{J.~O.} \bibnamefont{Sturm}}, \bibnamefont{and}
  \bibinfo{author}{\bibfnamefont{L.~M.~A.} \bibnamefont{Bettencourt}},
  \bibinfo{journal}{Sci. Adv.} \textbf{\bibinfo{volume}{1}},
  \bibinfo{pages}{e1400066} (\bibinfo{year}{2015}).

\bibitem[{\citenamefont{Milo et~al.}(2004{\natexlab{a}})\citenamefont{Milo,
  Itzkovitz, Kashtan, Levitt, Shen-Orr, Ayzenshtat, Sheffer, and
  Alon}}]{Milo-Itzkovitz-Kashtan-Levitt-ShenOrr-Ayzenshtat-Sheffer-Alon-2004-Science}
\bibinfo{author}{\bibfnamefont{R.}~\bibnamefont{Milo}},
  \bibinfo{author}{\bibfnamefont{S.}~\bibnamefont{Itzkovitz}},
  \bibinfo{author}{\bibfnamefont{N.}~\bibnamefont{Kashtan}},
  \bibinfo{author}{\bibfnamefont{R.}~\bibnamefont{Levitt}},
  \bibinfo{author}{\bibfnamefont{S.}~\bibnamefont{Shen-Orr}},
  \bibinfo{author}{\bibfnamefont{I.}~\bibnamefont{Ayzenshtat}},
  \bibinfo{author}{\bibfnamefont{M.}~\bibnamefont{Sheffer}}, \bibnamefont{and}
  \bibinfo{author}{\bibfnamefont{U.}~\bibnamefont{Alon}},
  \bibinfo{journal}{Science} \textbf{\bibinfo{volume}{303}},
  \bibinfo{pages}{1538} (\bibinfo{year}{2004}{\natexlab{a}}).

\bibitem[{\citenamefont{Milo et~al.}(2002)\citenamefont{Milo, Shen-Orr,
  Itzkovitz, Kashtan, Chklovskii, and
  Alon}}]{Milo-ShenOrr-Itzkovitz-Kashtan-Chklovskii-Alon-2002-Science}
\bibinfo{author}{\bibfnamefont{R.}~\bibnamefont{Milo}},
  \bibinfo{author}{\bibfnamefont{S.}~\bibnamefont{Shen-Orr}},
  \bibinfo{author}{\bibfnamefont{S.}~\bibnamefont{Itzkovitz}},
  \bibinfo{author}{\bibfnamefont{N.}~\bibnamefont{Kashtan}},
  \bibinfo{author}{\bibfnamefont{D.}~\bibnamefont{Chklovskii}},
  \bibnamefont{and} \bibinfo{author}{\bibfnamefont{U.}~\bibnamefont{Alon}},
  \bibinfo{journal}{Science} \textbf{\bibinfo{volume}{298}},
  \bibinfo{pages}{824} (\bibinfo{year}{2002}).

\bibitem[{\citenamefont{Milo et~al.}(2004{\natexlab{b}})\citenamefont{Milo,
  Kashtan, Itzkovitz, Newman, and
  Alon}}]{Milo-Kashtan-Itzkovitz-Newman-Alon-2004-XXX}
\bibinfo{author}{\bibfnamefont{R.}~\bibnamefont{Milo}},
  \bibinfo{author}{\bibfnamefont{N.}~\bibnamefont{Kashtan}},
  \bibinfo{author}{\bibfnamefont{S.}~\bibnamefont{Itzkovitz}},
  \bibinfo{author}{\bibfnamefont{M.~E.~J.} \bibnamefont{Newman}},
  \bibnamefont{and} \bibinfo{author}{\bibfnamefont{U.}~\bibnamefont{Alon}}
  (\bibinfo{year}{2004}{\natexlab{b}}),
  \bibinfo{note}{http://arxiv.org/abs/cond-mat/0312028}.

\bibitem[{\citenamefont{Kovanen et~al.}(2013)\citenamefont{Kovanen, Kaski,
  Kert{\'{e}}sz, and
  Saram{\"{a}}ki}}]{Kovanen-Kaski-Kertesz-Saramaki-2013-PNAS}
\bibinfo{author}{\bibfnamefont{L.}~\bibnamefont{Kovanen}},
  \bibinfo{author}{\bibfnamefont{K.}~\bibnamefont{Kaski}},
  \bibinfo{author}{\bibfnamefont{J.}~\bibnamefont{Kert{\'{e}}sz}},
  \bibnamefont{and}
  \bibinfo{author}{\bibfnamefont{J.}~\bibnamefont{Saram{\"{a}}ki}},
  \bibinfo{journal}{Proc. Natl. Acad. Sci. U.S.A.}
  \textbf{\bibinfo{volume}{110}}, \bibinfo{pages}{18070}
  (\bibinfo{year}{2013}).

\bibitem[{\citenamefont{Klimek and Thurner}(2013)}]{Klimek-Thurner-2013-NJP}
\bibinfo{author}{\bibfnamefont{P.}~\bibnamefont{Klimek}} \bibnamefont{and}
  \bibinfo{author}{\bibfnamefont{S.}~\bibnamefont{Thurner}},
  \bibinfo{journal}{New J. Phys.} \textbf{\bibinfo{volume}{15}},
  \bibinfo{pages}{063008} (\bibinfo{year}{2013}).

\bibitem[{\citenamefont{Ball and Newman}(2013)}]{Ball-Newman-2013-NS}
\bibinfo{author}{\bibfnamefont{B.}~\bibnamefont{Ball}} \bibnamefont{and}
  \bibinfo{author}{\bibfnamefont{M.~E.~J.} \bibnamefont{Newman}},
  \bibinfo{journal}{Network Science} \textbf{\bibinfo{volume}{1}},
  \bibinfo{pages}{16} (\bibinfo{year}{2013}).

\bibitem[{\citenamefont{Palla et~al.}(2007)\citenamefont{Palla, Barab{\'a}si,
  and Vicsek}}]{Palla-Barabasi-Vicsek-2007-Nature}
\bibinfo{author}{\bibfnamefont{G.}~\bibnamefont{Palla}},
  \bibinfo{author}{\bibfnamefont{A.-L.} \bibnamefont{Barab{\'a}si}},
  \bibnamefont{and} \bibinfo{author}{\bibfnamefont{T.}~\bibnamefont{Vicsek}},
  \bibinfo{journal}{Nature} \textbf{\bibinfo{volume}{446}},
  \bibinfo{pages}{664} (\bibinfo{year}{2007}).

\bibitem[{\citenamefont{Onnela et~al.}(2007)\citenamefont{Onnela, Saram{\"a}ki,
  Hyv{\"o}nen, Szab{\'o}, Lazer, Kaski, Kert{\'e}sz, and
  Barab{\'a}si}}]{Onnela-Saramaki-Hyvonen-Szabo-Lazer-Kaski-Kertesz-Barabasi-2007-PNAS}
\bibinfo{author}{\bibfnamefont{J.-P.} \bibnamefont{Onnela}},
  \bibinfo{author}{\bibfnamefont{J.}~\bibnamefont{Saram{\"a}ki}},
  \bibinfo{author}{\bibfnamefont{J.}~\bibnamefont{Hyv{\"o}nen}},
  \bibinfo{author}{\bibfnamefont{G.}~\bibnamefont{Szab{\'o}}},
  \bibinfo{author}{\bibfnamefont{D.}~\bibnamefont{Lazer}},
  \bibinfo{author}{\bibfnamefont{K.}~\bibnamefont{Kaski}},
  \bibinfo{author}{\bibfnamefont{J.}~\bibnamefont{Kert{\'e}sz}},
  \bibnamefont{and} \bibinfo{author}{\bibfnamefont{A.-L.}
  \bibnamefont{Barab{\'a}si}}, \bibinfo{journal}{Proc. Natl. Acad. Sci. U.S.A.}
  \textbf{\bibinfo{volume}{104}}, \bibinfo{pages}{7332} (\bibinfo{year}{2007}).

\bibitem[{\citenamefont{Kumpula et~al.}(2007)\citenamefont{Kumpula, Onnela,
  Saram{\"a}ki, Kaski, and
  Kert{\'e}sz}}]{Kumpula-Onnela-Saramaki-Kaski-Kertesz-2007-PRL}
\bibinfo{author}{\bibfnamefont{J.~M.} \bibnamefont{Kumpula}},
  \bibinfo{author}{\bibfnamefont{J.-P.} \bibnamefont{Onnela}},
  \bibinfo{author}{\bibfnamefont{J.}~\bibnamefont{Saram{\"a}ki}},
  \bibinfo{author}{\bibfnamefont{K.}~\bibnamefont{Kaski}}, \bibnamefont{and}
  \bibinfo{author}{\bibfnamefont{J.}~\bibnamefont{Kert{\'e}sz}},
  \bibinfo{journal}{Phys. Rev. Lett.} \textbf{\bibinfo{volume}{99}},
  \bibinfo{pages}{228701} (\bibinfo{year}{2007}).

\bibitem[{\citenamefont{Eagle et~al.}(2009)\citenamefont{Eagle, Pentland, and
  Lazer}}]{Eagle-Penland-Lazer-2009-PNAS}
\bibinfo{author}{\bibfnamefont{N.}~\bibnamefont{Eagle}},
  \bibinfo{author}{\bibfnamefont{A.}~\bibnamefont{Pentland}}, \bibnamefont{and}
  \bibinfo{author}{\bibfnamefont{D.}~\bibnamefont{Lazer}},
  \bibinfo{journal}{Proc. Natl. Acad. Sci. U.S.A.}
  \textbf{\bibinfo{volume}{106}}, \bibinfo{pages}{15274}
  (\bibinfo{year}{2009}).

\bibitem[{\citenamefont{Jo et~al.}(2011)\citenamefont{Jo, Pan, and
  Kaski}}]{Jo-Pan-Kaski-2011-PLoS1}
\bibinfo{author}{\bibfnamefont{H.-H.} \bibnamefont{Jo}},
  \bibinfo{author}{\bibfnamefont{R.~K.} \bibnamefont{Pan}}, \bibnamefont{and}
  \bibinfo{author}{\bibfnamefont{K.}~\bibnamefont{Kaski}},
  \bibinfo{journal}{PLoS One} \textbf{\bibinfo{volume}{6}},
  \bibinfo{pages}{e22687} (\bibinfo{year}{2011}).

\bibitem[{\citenamefont{Jiang et~al.}(2013)\citenamefont{Jiang, Xie, Li,
  Podobnik, Zhou, and Stanley}}]{Jiang-Xie-Li-Podobnik-Zhou-Stanley-2013-PNAS}
\bibinfo{author}{\bibfnamefont{Z.-Q.} \bibnamefont{Jiang}},
  \bibinfo{author}{\bibfnamefont{W.-J.} \bibnamefont{Xie}},
  \bibinfo{author}{\bibfnamefont{M.-X.} \bibnamefont{Li}},
  \bibinfo{author}{\bibfnamefont{B.}~\bibnamefont{Podobnik}},
  \bibinfo{author}{\bibfnamefont{W.-X.} \bibnamefont{Zhou}}, \bibnamefont{and}
  \bibinfo{author}{\bibfnamefont{H.~E.} \bibnamefont{Stanley}},
  \bibinfo{journal}{Proc. Natl. Acad. Sci. U.S.A.}
  \textbf{\bibinfo{volume}{110}}, \bibinfo{pages}{1600} (\bibinfo{year}{2013}).

\bibitem[{\citenamefont{Currarini et~al.}(2009)\citenamefont{Currarini,
  Jackson, and Pin}}]{Currarini-Jackson-Pin-2009-Em}
\bibinfo{author}{\bibfnamefont{S.}~\bibnamefont{Currarini}},
  \bibinfo{author}{\bibfnamefont{M.~O.} \bibnamefont{Jackson}},
  \bibnamefont{and} \bibinfo{author}{\bibfnamefont{P.}~\bibnamefont{Pin}},
  \bibinfo{journal}{Econometrica} \textbf{\bibinfo{volume}{77}},
  \bibinfo{pages}{1003} (\bibinfo{year}{2009}).

\bibitem[{\citenamefont{Currarini et~al.}(2010)\citenamefont{Currarini,
  Jackson, and Pin}}]{Currarini-Jackson-Pin-2010-PNAS}
\bibinfo{author}{\bibfnamefont{S.}~\bibnamefont{Currarini}},
  \bibinfo{author}{\bibfnamefont{M.~O.} \bibnamefont{Jackson}},
  \bibnamefont{and} \bibinfo{author}{\bibfnamefont{P.}~\bibnamefont{Pin}},
  \bibinfo{journal}{Proc. Natl. Acad. Sci. U.S.A.}
  \textbf{\bibinfo{volume}{107}}, \bibinfo{pages}{4857} (\bibinfo{year}{2010}).

\bibitem[{\citenamefont{Uzzi}(1996)}]{Uzzi-1996-ASR}
\bibinfo{author}{\bibfnamefont{B.}~\bibnamefont{Uzzi}},
  \bibinfo{journal}{American Sociological Review}
  \textbf{\bibinfo{volume}{61}}, \bibinfo{pages}{674} (\bibinfo{year}{1996}).

\bibitem[{\citenamefont{Guimera et~al.}(2014)\citenamefont{Guimera, Uzzi,
  Spiro, and Amaral}}]{Guimera-Uzzi-Spiro-Amaral-2005-Science}
\bibinfo{author}{\bibfnamefont{R.}~\bibnamefont{Guimera}},
  \bibinfo{author}{\bibfnamefont{B.}~\bibnamefont{Uzzi}},
  \bibinfo{author}{\bibfnamefont{J.}~\bibnamefont{Spiro}}, \bibnamefont{and}
  \bibinfo{author}{\bibfnamefont{L.-A.-N.} \bibnamefont{Amaral}},
  \bibinfo{journal}{Science} \textbf{\bibinfo{volume}{308}},
  \bibinfo{pages}{697} (\bibinfo{year}{2014}).

\bibitem[{\citenamefont{Cantner and Joel}(2011)}]{Cantner-Joel-2011-IUPJKM}
\bibinfo{author}{\bibfnamefont{U.}~\bibnamefont{Cantner}} \bibnamefont{and}
  \bibinfo{author}{\bibfnamefont{K.}~\bibnamefont{Joel}}, \bibinfo{journal}{IUP
  Journal of Knowledge Management} \textbf{\bibinfo{volume}{9}},
  \bibinfo{pages}{57} (\bibinfo{year}{2011}).

\bibitem[{\citenamefont{Garas et~al.}(2014)\citenamefont{Garas, Tomasello, and
  Schweitzer}}]{Garas-Tomasello-Schweitzer-2014-ArXiv}
\bibinfo{author}{\bibfnamefont{A.}~\bibnamefont{Garas}},
  \bibinfo{author}{\bibfnamefont{M.-V.} \bibnamefont{Tomasello}},
  \bibnamefont{and}
  \bibinfo{author}{\bibfnamefont{F.}~\bibnamefont{Schweitzer}},
  \bibinfo{journal}{arXiv preprint arXiv:1403.3298}  (\bibinfo{year}{2014}).

\bibitem[{\citenamefont{Jiang et~al.}(2009)\citenamefont{Jiang, Zhou, and
  Tan}}]{Jiang-Zhou-Tan-2009-EPL}
\bibinfo{author}{\bibfnamefont{Z.-Q.} \bibnamefont{Jiang}},
  \bibinfo{author}{\bibfnamefont{W.-X.} \bibnamefont{Zhou}}, \bibnamefont{and}
  \bibinfo{author}{\bibfnamefont{Q.-Z.} \bibnamefont{Tan}},
  \bibinfo{journal}{EPL (Europhys. Lett.)} \textbf{\bibinfo{volume}{88}},
  \bibinfo{pages}{48007} (\bibinfo{year}{2009}).

\bibitem[{\citenamefont{Jiang et~al.}(2010)\citenamefont{Jiang, Ren, Gu, Tan,
  and Zhou}}]{Jiang-Ren-Gu-Tan-Zhou-2010-PA}
\bibinfo{author}{\bibfnamefont{Z.-Q.} \bibnamefont{Jiang}},
  \bibinfo{author}{\bibfnamefont{F.}~\bibnamefont{Ren}},
  \bibinfo{author}{\bibfnamefont{G.-F.} \bibnamefont{Gu}},
  \bibinfo{author}{\bibfnamefont{Q.-Z.} \bibnamefont{Tan}}, \bibnamefont{and}
  \bibinfo{author}{\bibfnamefont{W.-X.} \bibnamefont{Zhou}},
  \bibinfo{journal}{Physica A} \textbf{\bibinfo{volume}{389}},
  \bibinfo{pages}{807} (\bibinfo{year}{2010}).

\bibitem[{\citenamefont{Thurner et~al.}(2012)\citenamefont{Thurner, Szell, and
  Sinatra}}]{Thurner-Szell-Sinatra-2012-PLoS1}
\bibinfo{author}{\bibfnamefont{S.}~\bibnamefont{Thurner}},
  \bibinfo{author}{\bibfnamefont{M.}~\bibnamefont{Szell}}, \bibnamefont{and}
  \bibinfo{author}{\bibfnamefont{R.}~\bibnamefont{Sinatra}},
  \bibinfo{journal}{PLoS One} \textbf{\bibinfo{volume}{7}},
  \bibinfo{pages}{e29796} (\bibinfo{year}{2012}).

\bibitem[{\citenamefont{Szell et~al.}(2012)\citenamefont{Szell, Sinatra, Petri,
  Thurner, and Latora}}]{Szell-Sinatra-Petri-Thurner-Latora-2012-SR}
\bibinfo{author}{\bibfnamefont{M.}~\bibnamefont{Szell}},
  \bibinfo{author}{\bibfnamefont{R.}~\bibnamefont{Sinatra}},
  \bibinfo{author}{\bibfnamefont{G.}~\bibnamefont{Petri}},
  \bibinfo{author}{\bibfnamefont{S.}~\bibnamefont{Thurner}}, \bibnamefont{and}
  \bibinfo{author}{\bibfnamefont{V.}~\bibnamefont{Latora}},
  \bibinfo{journal}{Sci. Rep.} \textbf{\bibinfo{volume}{2}},
  \bibinfo{pages}{457} (\bibinfo{year}{2012}).

\bibitem[{\citenamefont{Szell and Thurner}(2012)}]{Szell-Thurner-2012-ACS}
\bibinfo{author}{\bibfnamefont{M.}~\bibnamefont{Szell}} \bibnamefont{and}
  \bibinfo{author}{\bibfnamefont{S.}~\bibnamefont{Thurner}},
  \bibinfo{journal}{Adv. Complex Sys.} \textbf{\bibinfo{volume}{15}},
  \bibinfo{pages}{1250064} (\bibinfo{year}{2012}).

\bibitem[{\citenamefont{Bainbridge}(2007)}]{Bainbridge-2007-Science}
\bibinfo{author}{\bibfnamefont{W.~S.} \bibnamefont{Bainbridge}},
  \bibinfo{journal}{Science} \textbf{\bibinfo{volume}{317}},
  \bibinfo{pages}{472} (\bibinfo{year}{2007}).

\bibitem[{\citenamefont{Papagiannidis et~al.}(2008)\citenamefont{Papagiannidis,
  Bourlakis, and Li}}]{Papagiannidis-Bourlakis-Li-2008-TFSC}
\bibinfo{author}{\bibfnamefont{S.}~\bibnamefont{Papagiannidis}},
  \bibinfo{author}{\bibfnamefont{M.}~\bibnamefont{Bourlakis}},
  \bibnamefont{and} \bibinfo{author}{\bibfnamefont{F.}~\bibnamefont{Li}},
  \bibinfo{journal}{Tech. Forcast. Soc. Change} \textbf{\bibinfo{volume}{75}},
  \bibinfo{pages}{610} (\bibinfo{year}{2008}).

\bibitem[{\citenamefont{Williams}(2010)}]{Williams-2010-CT}
\bibinfo{author}{\bibfnamefont{D.}~\bibnamefont{Williams}},
  \bibinfo{journal}{Commun. Theory} \textbf{\bibinfo{volume}{20}},
  \bibinfo{pages}{451} (\bibinfo{year}{2010}).

\bibitem[{\citenamefont{Chesney et~al.}(2009)\citenamefont{Chesney, Chuah, and
  Hoffmann}}]{Chesney-Chuah-Hoffmann-2009-JEBO}
\bibinfo{author}{\bibfnamefont{T.}~\bibnamefont{Chesney}},
  \bibinfo{author}{\bibfnamefont{S.-H.} \bibnamefont{Chuah}}, \bibnamefont{and}
  \bibinfo{author}{\bibfnamefont{R.}~\bibnamefont{Hoffmann}},
  \bibinfo{journal}{J. Econ. Behav. Org.} \textbf{\bibinfo{volume}{72}},
  \bibinfo{pages}{618} (\bibinfo{year}{2009}).

\bibitem[{\citenamefont{Szell et~al.}(2010)\citenamefont{Szell, Lambiotte, and
  Thurner}}]{Szell-Lambiotte-Thurner-2010-PNAS}
\bibinfo{author}{\bibfnamefont{M.}~\bibnamefont{Szell}},
  \bibinfo{author}{\bibfnamefont{R.}~\bibnamefont{Lambiotte}},
  \bibnamefont{and} \bibinfo{author}{\bibfnamefont{S.}~\bibnamefont{Thurner}},
  \bibinfo{journal}{Proc. Natl. Acad. Sci. U.S.A.}
  \textbf{\bibinfo{volume}{107}}, \bibinfo{pages}{13636}
  (\bibinfo{year}{2010}).

\bibitem[{\citenamefont{Szell and Thurner}(2010)}]{Szell-Thurner-2010-SN}
\bibinfo{author}{\bibfnamefont{M.}~\bibnamefont{Szell}} \bibnamefont{and}
  \bibinfo{author}{\bibfnamefont{S.}~\bibnamefont{Thurner}},
  \bibinfo{journal}{Soc. Networks} \textbf{\bibinfo{volume}{32}},
  \bibinfo{pages}{313} (\bibinfo{year}{2010}).

\bibitem[{\citenamefont{Szell and Thurner}(2013)}]{Szell-Thurner-2013-SR}
\bibinfo{author}{\bibfnamefont{M.}~\bibnamefont{Szell}} \bibnamefont{and}
  \bibinfo{author}{\bibfnamefont{S.}~\bibnamefont{Thurner}},
  \bibinfo{journal}{Sci. Rep.} \textbf{\bibinfo{volume}{3}},
  \bibinfo{pages}{1214} (\bibinfo{year}{2013}).

\bibitem[{\citenamefont{Grabowski and
  Kosi{\'n}ski}(2008)}]{Grabowski-Kosinski-2008-APPA}
\bibinfo{author}{\bibfnamefont{A.}~\bibnamefont{Grabowski}} \bibnamefont{and}
  \bibinfo{author}{\bibfnamefont{R.}~\bibnamefont{Kosi{\'n}ski}},
  \bibinfo{journal}{Acta Phys. Pol. A} \textbf{\bibinfo{volume}{114}},
  \bibinfo{pages}{589} (\bibinfo{year}{2008}).

\bibitem[{\citenamefont{Grabowski and
  Kruszewska}(2007)}]{Grabowski-Kruszewska-2007-IJMPC}
\bibinfo{author}{\bibfnamefont{A.}~\bibnamefont{Grabowski}} \bibnamefont{and}
  \bibinfo{author}{\bibfnamefont{N.}~\bibnamefont{Kruszewska}},
  \bibinfo{journal}{Int. J. Mod. Phys. C} \textbf{\bibinfo{volume}{18}},
  \bibinfo{pages}{1527} (\bibinfo{year}{2007}).

\bibitem[{\citenamefont{Xie et~al.}(2016)\citenamefont{Xie, Li, Jiang, Tan,
  Podobnik, Zhou, and
  Stanley}}]{Xie-Li-Jiang-Tan-Podobnik-Zhou-Stanley-2016-SR}
\bibinfo{author}{\bibfnamefont{W.-J.} \bibnamefont{Xie}},
  \bibinfo{author}{\bibfnamefont{M.-X.} \bibnamefont{Li}},
  \bibinfo{author}{\bibfnamefont{Z.-Q.} \bibnamefont{Jiang}},
  \bibinfo{author}{\bibfnamefont{Q.-Z.} \bibnamefont{Tan}},
  \bibinfo{author}{\bibfnamefont{B.}~\bibnamefont{Podobnik}},
  \bibinfo{author}{\bibfnamefont{W.-X.} \bibnamefont{Zhou}}, \bibnamefont{and}
  \bibinfo{author}{\bibfnamefont{H.~E.} \bibnamefont{Stanley}},
  \bibinfo{journal}{Sci. Rep.} \textbf{\bibinfo{volume}{6}},
  \bibinfo{pages}{18727} (\bibinfo{year}{2016}).

\bibitem[{\citenamefont{Serrano et~al.}(2009)\citenamefont{Serrano,
  Bogu{\~{n}}{\'a}, and Vespignani}}]{Serrano-Boguna-Vespignani-2009-PNAS}
\bibinfo{author}{\bibfnamefont{M.~{\'A}.} \bibnamefont{Serrano}},
  \bibinfo{author}{\bibfnamefont{M.}~\bibnamefont{Bogu{\~{n}}{\'a}}},
  \bibnamefont{and}
  \bibinfo{author}{\bibfnamefont{A.}~\bibnamefont{Vespignani}},
  \bibinfo{journal}{Proc. Natl. Acad. Sci. U.S.A.}
  \textbf{\bibinfo{volume}{106}}, \bibinfo{pages}{6483} (\bibinfo{year}{2009}).

\bibitem[{\citenamefont{Stouffer et~al.}(2012)\citenamefont{Stouffer,
  Sales-Pardo, Sirer, and
  Bascompte}}]{Stouffer-SalesPardo-Sirer-Bascompte-2012-Science}
\bibinfo{author}{\bibfnamefont{D.}~\bibnamefont{Stouffer}},
  \bibinfo{author}{\bibfnamefont{M.}~\bibnamefont{Sales-Pardo}},
  \bibinfo{author}{\bibfnamefont{M.}~\bibnamefont{Sirer}}, \bibnamefont{and}
  \bibinfo{author}{\bibfnamefont{J.}~\bibnamefont{Bascompte}},
  \bibinfo{journal}{Science} \textbf{\bibinfo{volume}{335}},
  \bibinfo{pages}{1489} (\bibinfo{year}{2012}).

\bibitem[{\citenamefont{Burt}(2009)}]{Burt-2009}
\bibinfo{author}{\bibfnamefont{R.~S.} \bibnamefont{Burt}},
  \emph{\bibinfo{title}{{Structural holes: The social structure of
  competition}}} (\bibinfo{publisher}{Harvard university press},
  \bibinfo{year}{2009}).

\bibitem[{\citenamefont{Fucha and Thurner}(2014)}]{Fuchs-Thurner-2014-PLoS1}
\bibinfo{author}{\bibfnamefont{B.}~\bibnamefont{Fucha}} \bibnamefont{and}
  \bibinfo{author}{\bibfnamefont{S.}~\bibnamefont{Thurner}},
  \bibinfo{journal}{PLoS One} \textbf{\bibinfo{volume}{9}},
  \bibinfo{pages}{e103503} (\bibinfo{year}{2014}).

\bibitem[{\citenamefont{Yee et~al.}(2007)\citenamefont{Yee, Bailenson, Urbanek,
  F., and Merget}}]{Yee-Bailenson-Urbanek-Chang-Merget-2007-CPB}
\bibinfo{author}{\bibfnamefont{N.}~\bibnamefont{Yee}},
  \bibinfo{author}{\bibfnamefont{J.~N.} \bibnamefont{Bailenson}},
  \bibinfo{author}{\bibfnamefont{M.}~\bibnamefont{Urbanek}},
  \bibinfo{author}{\bibfnamefont{C.}~\bibnamefont{F.}}, \bibnamefont{and}
  \bibinfo{author}{\bibfnamefont{D.}~\bibnamefont{Merget}},
  \bibinfo{journal}{CyberPsychology \& Behavior} \textbf{\bibinfo{volume}{10}},
  \bibinfo{pages}{115} (\bibinfo{year}{2007}).

\end{thebibliography}
\end{document}